# ANOMALOUS BEHAVIOR OF THE DIFFUSION COEFFICIENT IN INTERACTING ADSORBATES

S. Zayzoune[1], M. Mazroui[1], Y. Boughaleb[1,2,3] and A. Kara[4]

[1] Université Hassan II- Mohammedia. Faculté des Sciences Ben M'sik.
Laboratoire de Physique de la Matière Condensée,
B.P. 7955, Casablanca, Morocco

[2] Université Chouaib Doukali, Faculté des Sciences. Laboratoire de Physique de la Matière Condensée, El Jadida, Morocco

[3] Member of Hassan II Academy of Sciences and Technology, Morocco

[4] Department of Physics, University of Central Florida, Orlando, FL 32816, USA

**Abstract**: Langevin simulations provide an effective way to study collective effects of Brownian particles immersed in a two-dimensional periodic potential. In this paper, we concentrate essentially on the behaviour of the tracer ($D_{Tr}$) and bulk ($D_B$) diffusion coefficients as function of friction ($\Gamma$). Our simulations show that in the high friction limit, the two physical quantities $D_{Tr}$ and $D_B$ present qualitatively the same behaviour, for both coupled and decoupled substrate potentials. However, for the low friction regime, and especially for the coupled potential case, an anomalous diffusion behaviour is found where $D_{Tr}\Gamma \sim \Gamma^{1-\sigma_{Tr}}$ and $D_B\Gamma \sim \Gamma^{1-\sigma_B}$, with $\sigma_{Tr} < \sigma_B < 1$. We also found that in the case of weak dynamical coupling between the ad-particles and the substrate, the exponents are not universal and rather depend on the potentials. Moreover, changes in the inter-particle potentials may reverse the behaviour to a normal one.

# I. INTRODUCTION

The diffusion properties of systems consisting of Brownian particles (atoms or ions) in an external periodic potential have been the subject of intensive research [1-8]. In spite of the considerable efforts, it continues to attract continuous interest and activity even though a century has passed since the appearance of the famous seminal work of Einstein on the subject [9]. The motion of atoms, vacancies, excitations, molecules and clusters of molecules on a variety of surfaces (crystalline, disordered or strained) is an active area of research due to its technological importance: Josephson junctions [7,8], superionic conductors [12], adsorbates on crystal surfaces [13], colloidal spheres [14] and polymers diffusing at interfaces [15] among many others [16]. The models describing these phenomena are of particular interest from the point of view of modern statistical mechanics.

The problem of single Brownian particle in a periodic potential is well understood for a wide range of of friction, both in one and two-dimensional systems. This problem was well described using different techniques/models, from stochastic equations, such as the Langevin equation (LE) and the Bhatnagar -Gross- Krook (BGK) equation [17,18]. The later has been applied to the cases involving an external periodic force field resulting from an adiabatic periodic potential [19-22]; to molecular dynamics, path integration and also to chaotic Hamiltonian dynamics [22-27]. Recently, renewed efforts have been devoted to the study of diffusion in coupled two-dimensional potentials, but limited to those cases where the inter-particle interactions were ignored. A first example of such studies worth mentioning is the work performed Chen et al. [28], where a Langevin simulation was performed for a *2D* potential of centred-rectangular symmetry. The second, performed by Ferrando et al. [29], involved solving numerically the Fokker–Planck equation, for which a solution was obtained by extending the matrix-continued-fraction method (MCFM) to a *2D* potential of square symmetry in which no energy barrier is present along the straight line adjoining the minima. These studies found that an *anomalous* dependence of the diffusion coefficient $D$ on the friction $\Gamma$ holds in the low friction regime with $D\Gamma \propto \Gamma^{1-\sigma}$ where *σ=0.5* in the first study [28] and *σ <1* in the second [29].

In contrast to a non-interacting particles system, approximations for strongly interacting particles systems have only been developed for limited special cases [30, 31] and a general understanding of the interaction effects is still lacking. In these studies, the inter-particle interaction is not described explicitly, but rather it was postulated as site exclusion (a particle cannot jump over an occupied site. Since this leads to a density-independent collective diffusion (in disagreement with experimental evidence), one has to introduce interaction between ad-particles, to obtain a density-dependent result. This question is of particular importance in understanding a variety of physical phenomena in condensed matter physics such as, chemical reactions, and adsorption/desorption, to name just two. As it has been pointed out, interactions between particles introduce new time scales, which lead to more complicated phenomena, and an analytical treatment of the Fokker-Planck equation is almost impossible [16].

The present contribution is an extension of our study [32] of Brownian motion in two-dimensional non-separable periodic potentials, and considering the case of interacting particles.

Our primary interest is the investigation of the dynamical properties by calculating the bulk diffusion $D_B$ coefficient, which is the relevant quantity for studying commensurability effects. These calculations show that the variation of $D_{Tr}\Gamma$ and $D_B\Gamma$ as function of the friction present an anomalous behaviour (i.e. $D_{Tr}\Gamma \sim \Gamma^{1-\sigma_{Tr}}$ and $D_B\Gamma \sim \Gamma^{1-\sigma_B}$) in low friction regime for non-interacting *and* interacting particles with exponents ($\sigma_{Tr}$ and $\sigma_B$) that are not universal but depend on the parameters of the two potentials ($V_A$, $V_{int}$). We further believe that the general features of our results should be fairly general and applied to 2D transport, in real systems.

The outline of the paper is as follows. In section **II**, we describe briefly the model and the method used in this study. In section **III**, we present results concerning the tracer and the bulk diffusion coefficient for non-interacting and interacting particles immersed in two-dimensional periodic potential. Finally, in section **IV** we summarize the results obtained.

## II. MODEL AND METHOD OF NUMERICAL SIMULATION

An appropriate way to study the problem of interacting Brownian particles in two-dimensional periodic potential is to solve the Fokker-Planck equation for the distribution function in four-dimensional phase space. Nevertheless, the Fokker-Planck equations are very expensive to solve numerically, even for dynamical systems possessing only a very modest numbers of degrees of freedom. On the other hand, the Fokker-Planck equations can be solved using an equivalent Langevin simulation, which reduces the four-dimensional partial differential equations into a group of stochastic ordinary differential equations. Compared to the Fokker-Planck equations, stochastic differential equations are not difficult to solve, and with the advent of modern supercomputers, it is possible to tackle very large numbers of realisations in order to treat the problem for very wide range of values of the parameters of the systems of interest.

Consider *N* particles diffusing in a *2D* potential of square symmetry $V_A(x, y)$. Thus we can, to an excellent approximation, describe the particle motion that follows the Fokker-Plank equation, by an ordinary coupled Langevin equation (LE) for N particles, as follows:

$$m_i \frac{d^2 r_i}{dt^2} = -m_i \gamma v_i + K_i(r) + \zeta_i(t) \tag{1}$$

Here the suffix *i* is the label of a mobile particle with corresponding mass $m_i$, velocity $v_i$ at position vector $r_i$. The quantities $-m_i \gamma v_i$ ($\gamma$ is the damping coefficient) and $K_i$ are the deterministic forces acting on particle *i*. The frictional damping force is often treated just as a phenomenological parameter. We assume that $K_i$ is derived from both the periodic single-particle potential $V_A$ originating from the substrate and the pair potential $V_{int}$, whose explicit expression will be specified later in section III.

$$K_i(r) = -\nabla_i V_{tot}(r) \qquad (2)$$

with

$$V_{tot}(r) = \sum_{k=1}^{N} V_A(r_k) + \sum_{k<l}^{N} V_{int}(r_k - r_l) \qquad (3)$$

where $r$ denotes the configuration of the system in real space:

$$r = (r_1, r_2, r_3, ..., r_N). \qquad (4)$$

The damping coefficient $\gamma$ and the random force $\zeta_i(t)$ both arise from the interaction of the mobile particle with the thermal vibrations of the substrate and are related by the fluctuation-dissipation theorem [26]:

$$\begin{cases} <\zeta_i(t)> = 0 \\ <\zeta_i(t)\zeta_j(t')> = 2m\gamma k_B T \delta_{ij} \delta(t-t') \end{cases} \qquad (5)$$

With $<...>$ denoting a thermal average; $k_B$ and $T$ are the Bolzmann's constant and the environmental temperature, respectively. The Brownian dynamics algorithm as proposed by Beeman is based on the solution of the coupled stochastic equations for a system of $N$ particles. It defines an accurate procedure to calculate static and dynamic properties of systems of interacting Brownian particles. Initially, $N$ particles are placed randomly in a square lattice array enclosed in a square box of length $L$. The particles are then allowed to move according to Eq.(1) until equilibrium is reach (about 15,000 time steps). The time step is assumed to be short enough and was taken to be $3 \times 10^{-14}$s, which was found to avoid the large fluctuations introduced by larger time steps. Periodic boundary conditions simulating an infinite system were imposed. The range of the potential was taken to be half the length of the box so that the effect of each particle is counted only once. If a particle leaves the box through the right boundary, a new particle simultaneously enters the box from the left, conserving the density. The model system we consider here consists of an ensemble of 2D interacting particles in presence of a periodic surface potential $V_A$, which we treat as an isotropic two-dimensional substrate with a square symmetry,

$$V_A(x,y) = V_A(r) = G_1[cos(q_0 x) + cos(q_0 y)] + G_2 cos(q_0 x) cos(q_0 y) \qquad (6)$$

where $q_0 = \dfrac{2\pi}{a}$ is a reciprocal lattice vector and $a$ is the separation between two nearest minima. The minima of the potential are organized in a square lattice on the surface and in the centers of the cells, with potential maxima at the four corners, and saddle points at midpoints of the edges. The activation energy barriers are:

$$V_0 = 2(G_2 - G_1)  \quad (7)$$

For $G_1=G_2$ the minimum energy and the saddle energy coincide and give rise to channels of constant energy. The coupling term $G_2$ is responsible for the energy transfer between the *x* and *y* degrees of freedom and leads to qualitatively new dynamical features in the case of Hamiltonian systems. This type of potential represents a simple, but yet rich, 2D potential to investigate the effects of *x-y* coupling. This form has been largely used as a model periodic potential in theoretical studies of many different problems, such as the non-linear conservative dynamics of a classical particle [33], the collective diffusion of particles in super-ionic conductors [34] and the noise-activated diffusion of a classical particle [35]. When the interaction among particles is very weak, one can approximately treat the above problem in terms of Brownian motion of a single particle under a biased potential.

### III. RESULTS

It has been demonstrated previously that the transport and dynamic properties of particles on a surface can be significantly affected by inter-particles. In this context, Zhigang Zheng et al. [36] investigated the collective transport behaviour of an under-damped Frenkel-Kontorova chain, and found complex behaviour. Theirs results demonstrate clearly that the coupling between particles may enhance the diffusion process, depending on the inter-play between the harmonic chain and the substrate potential.

In the following, we present results on the diffusion coefficient as function of the friction, for both interacting and non-interacting cases. We recall that the following results are obtained from numerical simulations based on the solution of 2*N*-coupled Langevin equations (Eq.1) with periodic boundary conditions. The range of applications of this equation makes it a highly popular one, both for theoreticians and for experimentalists. Throughout the paper we use the dimensionless quantity $\Gamma = \frac{2\pi\gamma}{\omega_0}$ where $\omega_0 = q_0\sqrt{V_0/2m}$ is the vibration frequency of a single particle at the minimum of the substrate potential. This parameter is varied from one simulation to another. In contrast, one single value of the energy barrier is used in all simulations: $V_O = 0,1$ eV, which corresponds to 3.8*E*-3 in atomic unit (a.u). We have set $\frac{V_o}{k_BT} = 4$ reflecting that the temperature *T* remains constant also.

### III.1. Tracer diffusion coefficient: non-interacting particles

We consider here a system of ad-particles in equilibrium with the substrate. This is the only limit in which the diffusion coefficient of ad-particles can be defined accurately. The tracer diffusion coefficient $D_{Tr}$ is a quantity of fundamental importance to the diffusion of particles. It is defined as:

$$D_{Tr} = \frac{1}{2d} \lim_{t \to \infty} \frac{1}{t} \left\langle \frac{1}{N} \sum_i |r_i(t) - r_i(0)|^2 \right\rangle$$

$$= \frac{1}{d} \int_0^\infty \frac{1}{N} \sum_{i=1}^N \langle v_i(t) v_i(0) \rangle dt \qquad (9)$$

where the angle brackets $\langle \cdots \rangle$ means average over the initial conditions and over the stochastic trajectories obtained numerically by integrating the Langevin equation; $N$ is the number of particles, $d$ is the dimensionality of the space, $r_i(t)$ is the position vector of particle $i$ at time $t$, and $v_i(t)$ its velocity. The function $\phi(t) = \langle v_i(t) v_i(0) \rangle$ is the velocity-velocity correlation function associated with the tracer diffusion. This function is very interesting and readily accessible through STM experiments [37]. In order to assess the accuracy of the Langevin dynamic simulations, let us first begin with the study of the effects of $x$-$y$ coupling on the diffusion process corresponding to non-interacting particles ($V_{int}=0$). Essentially, we analyzed the situations when the $x$ and $y$ degrees of freedom are either coupled or decoupled.

Fig.1 shows the variation of $D_{Tr}\Gamma$ as a function of friction $\Gamma$ for separable and non-separable potentials. For the separable (decoupled) case $G_2$ (full square), $D_{Tr}\Gamma$ presents an inflexion point in the intermediate region (IF) and converges slowly to a constant value $D_{Tr} \sim 1/\Gamma$ at high (HF) friction regime. While for non-separable (coupled) case $G_2 \neq 0$, the tracer diffusion coefficient presents two different behaviours. At high friction regime (HF), the tracer diffusion coefficient presents the same behaviours as found for the decoupled case $D_{Tr} \sim 1/\Gamma$, while at low friction regime(HF), this quantity presents a linear behaviours $D_{Tr}\Gamma \sim \Gamma^{1-\sigma_{Tr}}$; with $\sigma_{Tr} < 1$. The absence of $x$-$y$ coupling has therefore the effect of making the diffusive motion fast. This substantial change in diffusion coefficient with the coupling term $G_2$ observed is easily understood: in the presence of the coupling, the width of the channel is narrower at the saddle point position than at the minimum. Consequently, the particles find it more difficult to overcome the barrier. Our Langevin dynamics simulations for these cases (separable and non-separable potential) recover the results obtained earlier by different researchers [28,29] using different techniques of calculations such as the matrix continued fraction (MCFM) and Molecular dynamics.

In order to extend this study to realistic systems, the interaction between particles must be introduced. Henceforth our main interest will be focused on the effects of the inter-particle interaction, which provide additional phenomena with respect to the simple case of non-interacting particles.

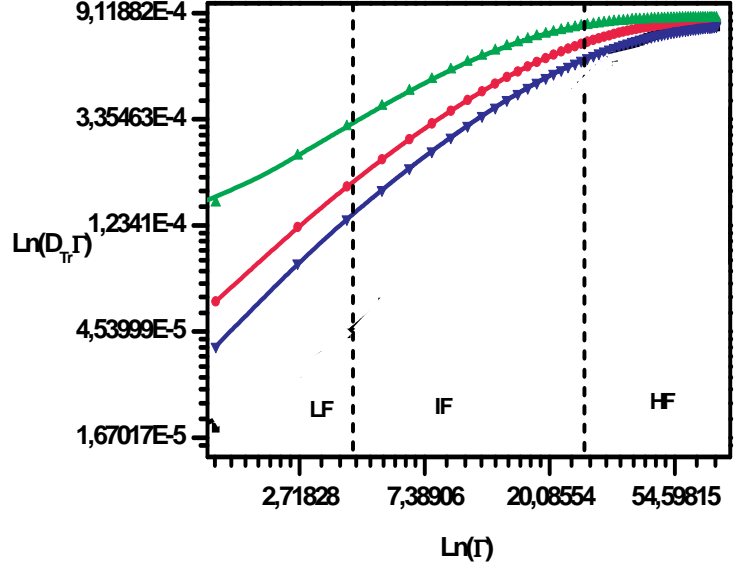

**Fig.1:** Variation of $D_{Tr}\Gamma$ as function of $\Gamma$ in decoupled case $G_2=0$ (full diamond) and coupled cases: $G_2=3E-6$ a.u (full circle), $G_2=7E-3$ u.a (full triangle). Three friction regimes are indicated: high friction (HF), intermediate friction (IF), and low friction (LF)

### III.2. Bulk diffusion coefficient: Effects of adparticle-adparticle interactions

In real surface systems, there are non-negligible adparticle-adparticle interactions which often dominate the behaviour of adsorbates dynamics at low temperatures [38]. When particle-particle interactions are present in the system, the diffusion process becomes correlated, and a distinction has to be made between single particle or tracer, and collective processes. In this important context of interacting particles ($V_{int}\neq 0$), an interesting problem is the behaviour of bulk diffusion coefficient $D_B$ [37], which is relevant for studying commensurability effects:

$$D_B = \frac{1}{2Ndt}\lim_{t\to\infty}\sum_{i,j}\left\langle (\vec{r}_i(t)-\vec{r}_j(0))^2 \right\rangle \\ = \frac{1}{d}\int_0^\infty \frac{1}{N}\sum_{i,j}\left\langle v_i(t)v_j(0)\right\rangle dt \qquad (10)$$

The parameters in this equation have the same meaning as defined above in Eq.(7). One may notice that the diffusion coefficient relates to the temperature, the interaction term and the concentration. For non-interacting Brownian particles $D_{Tr}=D_B$, since the velocity of any given particle is in no way depend on the other particles and the off diagonal terms $(i\neq j)$ in the expression for $D_B$ vanish.

The model system we consider here consists of an ensemble of 2D particles in presence of periodic surface potential. The particles interact with each other through the Yukawa pair potential:

$$V_{int}(r) = \frac{Q^2}{r}\exp(-\frac{r}{r_C}) \; ; \qquad r = \sqrt{x^2 + y^2} \qquad (11)$$

where $r$ is the separation between the particles and $Q$ is the effective charge. A cut-off at $r_C$ was used in the simulation, and is taken to be equal to $L/2$, where $L$ represents the length of the system.

In Fig.2, we have reported the variation of the bulk diffusion coefficient as function of friction, in decoupled case, for two values of the interaction term $Q^2$. The analysis of this figure show that $D_B$ presents an inflexion point in the intermediate regime (IF) and tends slowly to a constant $D_B \sim 1/\Gamma$ at low (LF) and high (HF) friction regime. We also note that $D_B \Gamma$ increases with the increasing of the interaction term.

While for the coupled case (see Fig.3), the shape of $D_B$ differs strongly from that computed in decoupled case, at high friction regime (HF) $D_B$ tends to a constant $D_B \sim \Gamma^{-1}$. While at low friction regime (LF) the bulk diffusion coefficient presents an anomalous behaviour $D_B \Gamma \sim \Gamma^{1-\sigma_B}$ with $\sigma_B < 1$ ($\sigma_B = 0.5$ at $Q^2 = 0.8$ and $\sigma_B = 0.1$ at $Q^2 = 0.7$). From the simulation results presented here, it becomes clear that the exponent $\sigma_B$ is not universal but depends also on the parameters of the pair potential. This expected result can be simply interpreted as a change in the structure of the periodic potential: the mutual interaction of the diffusing particles induces an effective potential which have a completely different shape from the substrate surface potential (Eq.6). These important results extend and confirm the earlier results of Chen et al. [28] and Caratti et al. [29]. On the other hand, our results indicate that if one introduces strong coupling strength between particles, then a higher diffusion rate can be achieved. In many realistic applications one hopes that the diffusion process can be improved effectively. Our exploratory work indicates that the coupling strengths may enhance the diffusion process which can be easily explained. If the interaction is very strong, it creates a quasi-regular arrangement of the particles by forcing some of them to move away from the minima. Such a situation, in which the substrate periodic potential is only a perturbation, affects significantly the transport properties. Consequently, the correlations between particles are then forward correlations and can enhance the diffusion process.

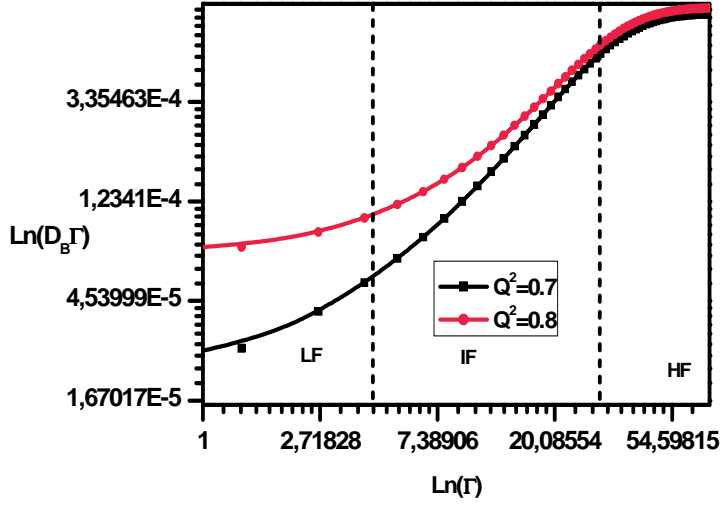

**Fig.2:** Variation of $D_B\Gamma$ as function of $\Gamma$ in decoupled in case $G_2=0$, for two values of the interaction term $Q^2$. The parameter $Q$ is expressed also in atomic unit

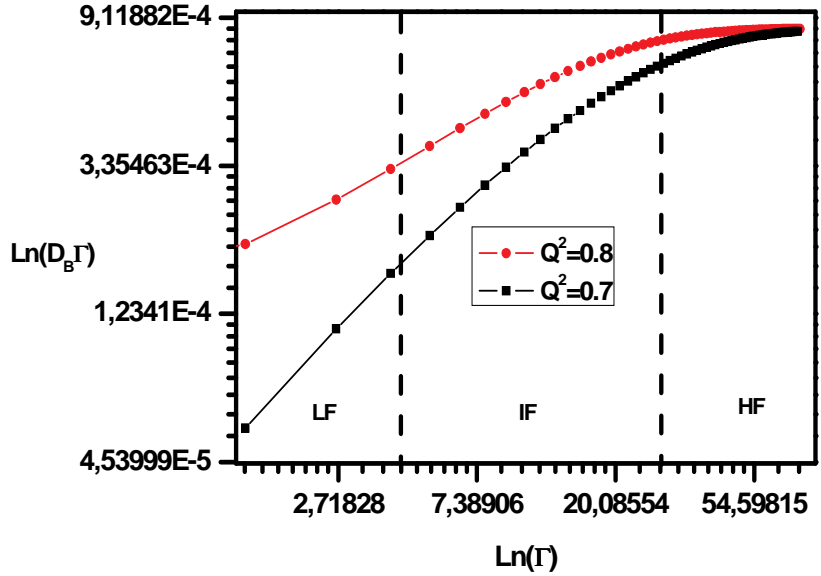

**Fig.3:** The same as Fig.2 but for coupled case $G_2=7E-3\ u.a$ for two values of the interaction term $Q^2$.

Finally, in order to determine how the exponents varied with the coupling term $G_2$, we have reported the variation of the exponents $\sigma_{Tr}$ and $\sigma_B$ as function of $G_2$ (Fig.4). The first phenomenon we observe is that the two relations between the coupling term $G_2$ and the exponents are monotonic, i.e., the exponents $\sigma_{Tr}$ and $\sigma_B$ decrease with increasing the coupling term. Secondly, our calculations show that the two exponents are equal only for $G_2=0$. So, the normal behaviour ($\sigma_B = \sigma_{Tr} =1$) can be achieved essentially in decoupled cases independently of the pair interaction. However, the inclusion of the interaction between particles change the exponent and tends to make it less anomalous ($\sigma_B > \sigma_{Tr}$), so that it is closer to a normal behaviour. Thus, we can confirm that these quantities are not universal but they depend on the parameters of the substrate potential and the pair interaction which keeps the system away from an anomalous situation.

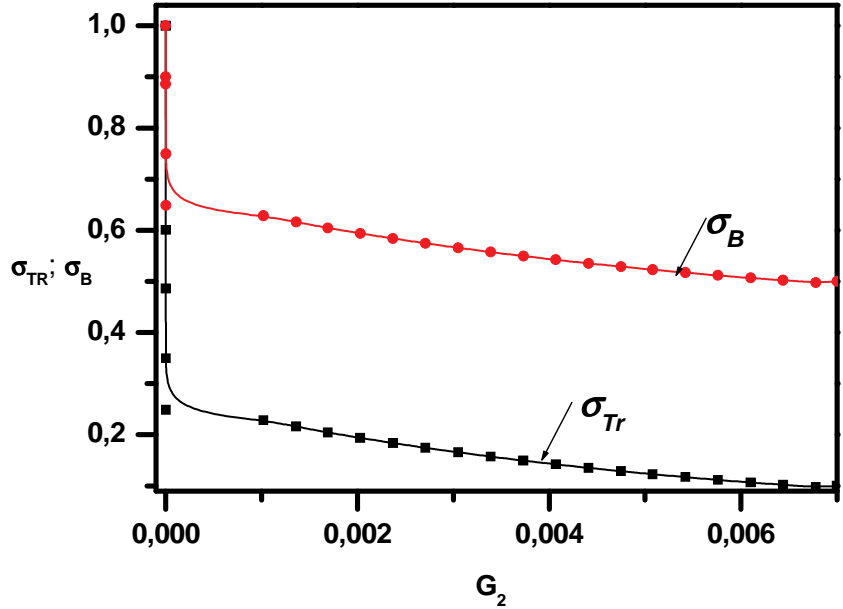

**Fig.4:** Variation of $\sigma$ as a function of the coupling term $G_2$ in the case of non-interacting (full square) and interacting adparticles with $Q^2=0.7$ (full diamond).

## IV. SUMMARY AND CONCLUSIONS

As a natural continuation of our earlier work dealing with application of simulational methods, we have studied in this paper the motion of Brownian particles immersed in a periodic potential, by means of the Langevin coupled equations. The method is general, and can be applied to systems of any arbitrary dimension. Here, two-dimensional systems have been considered, as models on crystal surfaces. The main goal of this study is to give the effects of the parameters of the periodic potential and the pair potential on the diffusion mechanisms at low friction (LF). We have focused our attention mainly on bulk diffusion $D_B$, and some results concerning single-particle diffusion have been given for comparison. It proves that due to an additional pair potential, new effects occur in the transport processes of Brownian particles, in particular if the periodic potential is coupled. In fact, for non-interacting particles, our Langevin dynamics simulations recover previously reported results ( $D_{Tr}\Gamma \sim \Gamma^{1-\sigma_{Tr}}$ with $\sigma_{Tr}$ <1) obtained by using different techniques of calculations such as the matrix continued fraction and Molecular dynamics.

On the other hand, in order to know the effects of the interaction between particles on diffusion, we have calculated the bulk diffusion coefficient for different values of the interacting term. The computations show that this quantity presents qualitatively the same behaviour as the tracer diffusion, namely that $D_B \sim 1/\Gamma$ for a decoupled case and a $D_B\Gamma \sim \Gamma^{1-\sigma_B}$ (at low friction regime) for a coupled case, with $\sigma_B$ <1. The exponents are not universal; they depend not only of the parameters of the substrate potential but also on the pair potential. Another important point worth mentioning is the enhancement in diffusion coefficient for the case of interacting particles. Thus we conclude that a higher diffusion process can be achieved in the case of strong coupled substrate potential and for strong strength of pair potentials. This information is quite useful as guidance for experimentalist searching for conditions under which they can control the diffusion regime.

**Acknowledgement**

The work of AK was partially supported by a UCF start up grant.